\documentclass[openacc]{rsproca_new}

\begin{document}

\title{Constancy of the Cluster Gas Mass Fraction in the $R_{\rm h}=ct$ Universe}

\author{F. Melia}

\address{Department of Physics, The Applied Math Program, and Department of Astronomy,
The University of Arizona, AZ 85721, USA}

\subject{Cosmology, Galaxies, Relativity}

\keywords{Baryon Fraction, Galaxy Clusters, Theoretical Cosmology}

\corres{F. Melia\\
\email{fmelia@email.arizona.edu}}

\begin{abstract}
The ratio of baryonic to dark matter densities is assumed to have remained 
constant throughout the formation of structure. With this, simulations 
show that the fraction $f_{\rm gas}(z)$ of baryonic mass to total mass in 
galaxy clusters should be nearly constant with redshift $z$. However,
the measurement of these quantities depends on the angular distance to the source,
which evolves with $z$ according to the assumed background cosmology. An accurate 
determination of $f_{\rm gas}(z)$ for a large sample of hot ($kT_e> 5$ keV), 
dynamically relaxed clusters could therefore be used as a probe of the 
cosmological expansion up to $z< 2$. The fraction $f_{\rm gas}(z)$
would remain constant only when the ``correct" cosmology is used to fit the data. 
In this paper, we compare the predicted gas mass fractions for both $\Lambda$CDM 
and the $R_{\rm h}=ct$ Universe and test them against the 3 largest cluster samples 
\cite{1,2,3}. We show that $R_{\rm h}=ct$ is consistent with a constant
$f_{\rm gas}$ in the redshift range $z\lesssim 2$, as was previously shown 
for the reference $\Lambda$CDM model (with parameter values $H_0=70$ km s$^{-1}$ 
Mpc$^{-1}$, $\Omega_m=0.3$ and $w_\Lambda=-1$). Unlike $\Lambda$CDM, however, 
the $R_{\rm h}=ct$ Universe has no free parameters to optimize in fitting the 
data. Model selection tools, such as the Akaike Information Criterion (AIC) 
and the Bayes Information Criterion (BIC), therefore tend to favor 
$R_{\rm h}=ct$ over $\Lambda$CDM. For example, the BIC favours $R_{\rm h}=ct$ 
with a likelihood of $\sim 95\%$ versus $\sim 5\%$ for $\Lambda$CDM.
\end{abstract}


\begin{fmtext}
\section{Introduction}
The idea that clusters might provide an independent probe of cosmological
expansion took root following a series of non-radiative hydrodynamical
simulations showing that the gas mass fraction, $f_{\rm gas}=M_{\rm gas}/
M_{\rm tot}$, in 
\end{fmtext}

\maketitle

\noindent the largest (i.e., $kT>5$ keV) dynamically relaxed clusters,
remains approximately constant with redshift \cite{4,5}. Here, $M_{\rm gas}$ 
is the mass of the intracluster medium and $M_{\rm tot}$ is the total mass of the 
cluster. These results followed seminal papers by Sasaki \cite{6} and Pen \cite{7},
who argued that the measurement of apparent evolution (or non-evolution) in
$f_{\rm gas}$ could be used to examine the angular distance to 
these sources as a function of redshift. Since then, several groups have 
pursued this line of work, compiling extensive catalogs of clusters suitable 
for such a study \cite{1,2,3,8,9}. In conjuction with this observational work,
more elaborate simulations, incorporating several key physical ingredients, such
as radiative cooling and the dynamical impact of turbulence, have provided a more
realistic assessment of the conditions and cluster size for which $f_{\rm gas}$
might in fact be expected to remain constant.

Under the (as yet unproven) assumption that the baryonic to dark
matter densities, $\rho_b/\rho_{d}$, is independent of redshift at
least out to $z\lesssim 2$, the geometry of the Universe can be constrained
in this way because the measured baryonic mass fraction depends on the
assumed angular diameter distance to the source, which is used along with
the inferred density to obtain $M_{\rm gas}$. (This assumption may have
to be modified if, and when, new physics beyond the standard model
implies that baryons and/or dark matter may be created or annihilated with
time since the big bang.) The baryonic matter content of galaxy clusters is
dominated by the X-ray emitting intracluster gas, whose mass exceeds that
of optically luminous material by a factor $\sim 6$ \cite{10,11}.
The other contributions to the total baryon
budget are expected to be very small. The emissivity of the X-ray
emitting gas is proportional to the square of its density, so the gas
mass profile in a cluster can be determined precisely from X-ray data.
Measuring $M_{\rm tot}$ is more difficult because it is often based on
the assumption of hydrostatic equilibrium in the gas, from which one
may infer the depth of the gravitational potential required to maintain
the density profile. Thus, only dynamically relaxed systems can be used
for this purpose. In addition, as we will consider in more detail below,
one would expect a constant mass fraction only for the very massive clusters,
with minimal impact from astrophysical factors, such as feedback and cooling.
But though cosmological simulations suggest that
under some circumstances $f_{\rm gas}$ should be invariant with redshift,
we would only see this in the data if the underlying model used to interpret
the measurements is the correct cosmology. So in principle one may
carry out a comparative test between competing models to see
which, if any, predicts a constant value of $f_{\rm gas}$ with
changing $z$.

\subsection{The Constancy of $f_{\rm gas}$}
The caveat here is that the constancy of $f_{\rm gas}$ with
redshift should be independent of the expansion dynamics. One
would certainly be justified in expecting this if $\rho_b/\rho_{d}$
has not changed with $z$. However,
to use this feature as a cosmological tool, we have to believe that
a sample of clusters exists for which $f_{\rm gas}$ is independent
of which version of $\Lambda$CDM (or other cosmology) we are
comparing with the data. For if the behavior of $f_{\rm gas}$
with redshift were different for different expansion rates,
we could not be certain that $f_{\rm gas}$ should in fact remain
constant.

In their high-resolution simulations, Kravtsov et al. \cite{12} incorporated
the effects of radiative cooling and galaxy formation on the baryon
fraction, including the impact on star formation, metal enrichment,
and stellar feedback. These processes increase the total baryon
fraction within scales as large as the virial radius, though it is
within the cluster cores that baryon fractions larger than the universal
value are seen. However, even with cooling, the cumulative baryon
fraction is close to its universal value at radii $r>r_{2,500}$, where
$r_\Delta$ is defined to be the radius within which the average cluster
density is greater than its critical value by a factor $\Delta$ at
that redshift (see equation~2.12 below). Moreover, even though the baryon
fraction may be different from its universal value at smaller radii,
simulations such as this suggest that the total baryon fraction within
the cluster virial radius does not evolve with time, regardless of
whether or not cooling is included.

Simulations with even greater sophistication than these were carried
out by Ettori et al. \cite{13}, this time including also the effects of
feedback through galactic winds and conduction. They found that the
baryon fraction within a fixed overdensity increases slightly with
redshift, though the impact at large cluster-centric distances
(i.e., $r>r_{500}$) is nearly independent of the physics included
in the calculations. More recently, Planelles et al. \cite{14} updated
these simulations by including feedback from supernovae and active
galactic nuclei. They too found that the baryon fraction is nearly
independent of the physical processes, and is characterized by a
negligible redshift evolution, if the cluster mass $M_{500}$ at
$r_{500}$ is $\gtrsim 10^{14}\;M_\odot$. At smaller radii, $r_{2,500}$,
its value slightly decreases, while its scatter increases by about
a factor of 2. As we shall see, the cluster catalogs currently
available for this cosmological test differ in their assumed overdensity
factor $\Delta$, so $f_{\rm gas}$ may not be uniformly constant with
redshift from sample to sample. This is an important caveat to
consider when weighing the results of model comparisons using
different cluster catalogs. In this paper, we will consider the
3 largest samples, two of which assume $\Delta=2,500$ \cite{1,2}, 
while the third adopts the value $\Delta=500$ \cite{3}.

\subsection{Testing Cosmological Models with $f_{\rm gas}$}
Up until now, the cluster gas mass fraction has been used
to probe only the parameter space associated with the standard
model of cosmology, $\Lambda$CDM. The reference model often
used for this work assumes a spatially flat universe ($k=0$) with a scaled
matter density $\Omega_m\equiv \rho_m/\rho_c=0.3$, where
$\rho_m=\rho_b+\rho_{d}$ is the matter density and $\rho_c\equiv 3c^2H_0^2/
8\pi G$ is the critical density in terms of the Hubble constant
$H_0$ today, and a dark energy in the form of
a cosmological constant with equation-of-state $w_{\Lambda}\equiv
p_{\Lambda}/\rho_{\Lambda}=-1$, in terms of its pressure $p_{\Lambda}$
and density $\rho_{\Lambda}$. Since these clusters lie at redshifts
$z< 1.5$, where the contribution of radiation to the total energy density
is below detectability, one can also assume that $\Omega_m+
\Omega_{\Lambda}=1$. In obvious notation, $\Omega_\Lambda\equiv
\rho_\Lambda/\rho_c$.

But in recent years, evidence has been accumulating that $\Lambda$CDM
is perhaps the empirical approximation to a more theoretically motivated FRW
cosmology known as the $R_{\rm h}=ct$ Universe \cite{15,16,17}.
(A somewhat pedagogical treament may be found in 
ref. \cite{18}.) The latter arises when
one invokes Birkhoff's theorem \cite{19} together with Weyl's
postulate \cite{20}, which lead to an identification of the Hubble
radius $R_{\rm h}=c/H$ as another manifestation of the Universe's
gravitational horizon, $2GM/c^2$, defined in terms of the
Misner-Sharp mass $M$ contained within a proper spherical volume
of radius $R_{\rm h}$ \cite{21}. It must therefore
itself be a proper distance $R_{\rm h}=a(t)r_{\rm h}$, where
$a(t)$ is the universal expansion factor and $r_{\rm h}$ is an
unchanging co-moving distance. This form of $R_{\rm h}$ leads
immediately to the condition that $\dot{a}=$ constant. Thus,
the $R_{\rm h}=ct$ Universe expands at a constant rate.

This cosmology should not be confused with the Milne Universe
\cite{22}, which is empty and has negative spatial curvature
($k=-1$). The Milne Universe does not at all fit the
cosmological data and was ruled out as a viable model
long ago. Instead, the $R_{\rm h}=ct$ Universe is flat ($k=0$)
and predicts very simple, analytical forms for measurable
quantities, such as the luminosity distance,
\begin{equation}
d_L^{R_{\rm h}}=R_{\rm h}(1+z)\ln(1+z)\;,
\end{equation}
and the redshift dependence of the Hubble constant,
\begin{equation}
H(z)=H_0(1+z)\;.
\end{equation}
We will provide a more detailed description of the differences
between $R_{\rm h}=ct$ and $\Lambda$CDM in \S~3 below.

By now, the predictions of these two cosmologies have been
compared to each other using many diverse tests
and available data, including: the cosmic chronometers
\cite{23}, the gamma-ray burst Hubble diagram
\cite{24}, the high-$z$ quasars \cite{25}, the angular
correlation function of the cosmic microwave background radiation
\cite{26}, and the high-$z$ galaxies \cite{27}, among
others (some not yet published). The consensus from all of
this work appears to be that the $R_{\rm h}=ct$ Universe
is closer to the correct cosmology than $\Lambda$CDM is.

In this paper, we extend this comparative study even further,
by now examining the role played by the $R_{\rm h }=ct$ Universe
in maintaining an approximately constant value of the cluster gas 
mass fraction in the redshift range $z\lesssim 2$.
In \S2 of this paper, we briefly describe the basic
theory behind this independent cosmological probe. In \S3, we
discuss the assumptions necessary in both
$\Lambda$CDM and $R_{\rm h}=ct$ to make this diagnostic
meaningful for cosmology, and then assemble the most extensive
catalogs now available in \S4. We carry out our direct comparison
between $\Lambda$CDM and $R_{\rm h}=ct$ in \S5, and then discuss
our results and place them in a proper context in \S6.

\section{The Use of Cluster Gas Mass Fraction as a Cosmological Probe}
The baryonic matter content of galaxy clusters is dominated by the
X-ray-emitting intracluster gas predominantly via thermal
bremsstrahlung \cite{28}. Thus, for the spherical $\beta$-model
profile \cite{29}, the gas mass $M_{\rm gas}(<R)$
within a radius $R$ derived from X-ray observations may be written
\begin{eqnarray}
M_{\rm gas}(<R)&=&\left[{3\pi\hbar m_ec^2\over 2(1+X)e^6}\right]^{1/2}
\left({3m_ec^2\over 2\pi k_B T_e}\right)^{1/4}m_H\times\nonumber\\
&\null&\qquad\qquad{1\over [\bar{g}_B(T_e)]^{1/2}}\,r_c^{3/2}\,\left[{I_M(R/r_c,\beta)\over
I_L^{1/2}(R/r_c,\beta)}\right]\left[L_X(<R)\right]^{1/2}\;,
\end{eqnarray}
where $m_e$ and $m_H$ are the electron and hydrogen masses, respectively,
$X$ is the hydrogen fraction by mass, $T_e$ is the (electron) gas temperature,
$\bar{g}_B(T_e)$ is the Gaunt factor, $r_c$ is the core radius, and
$I_M(y,\beta)\equiv \int_0^y (1+u^2)^{-3\beta/2}\,u^2\,du$,
$I_L(y,\beta)\equiv \int_0^y (1+u^2)^{-3\beta}\,u^2\,du$.
The chosen cosmology enters this expression in three ways: through
the X-ray luminosity
\begin{equation}
L_X(<R)=4\pi d_L^2\,f_X(<\theta)\;,
\end{equation}
through the core radius
\begin{equation}
r_c=\theta_c\,d_A\;,
\end{equation}
and through the variable radius
\begin{equation}
R=\theta\,d_A\;,
\end{equation}
in terms of the observed angular size $\theta$, the observed
X-ray flux $f_X$, and the luminosity ($d_L$) and angular ($d_A$) distances.
Some of the data we will examine below are based on observations of the
Sunyaev-Zeldovich effect (SZE), for which $M_{\rm gas}$ depends
on a different power of radius. For the SZE objects, $M_{\rm gas}
\propto d_A^2$, instead of $\propto d_L\,d_A^{3/2}$ \cite{1}, which
is applicable in all other cases:
\begin{equation}
M_{\rm gas}(z,<\theta)\propto d_L\,d_A^{3/2}\;.
\end{equation}
And since $d_A={(1+z)^{-2}d_L}$, we have
\begin{equation}
M_{\rm gas}(z,<\theta) \propto d_A^{5/2}
\end{equation}
at any given redshift $z$.

Under the assumption of hydrostatic equilibrium and isothermality ($T_e=$ constant), the
total mass within radius $R$ is given by
\begin{equation}
M_{\rm tot}(<R)=-\left.{k_B T_e\,R\over G\mu m_H}{{\rm d\,ln}\,n_e(r)\over{\rm d\,ln}\,r}\right|_{r=R}\;,
\end{equation}
where $\mu$ is the mean-molecular weight per particle and $n_e(r)$ is the spatially-dependent
electron number density. Thus, one gets
\begin{equation}
M_{\rm tot}(<\theta)\propto d_A\;,
\end{equation}
and therefore
\begin{equation}
f_{\rm gas}\propto d_A^{3/2}\;.
\end{equation}

According to Equation~(1.1), the angular distance in these expressions takes on a very simple analytical
form in the $R_{\rm h}=ct$ Universe:
\begin{equation}
d_A^{R_{\rm h}}={R_{\rm h}\over (1+z)}\ln(1+z)\;.
\end{equation}
The corresponding expression in $\Lambda$CDM is
\begin{equation}
d_{A}^{\Lambda}={R_{\rm h}\over(1+z)}{1\over\sqrt{\mid\Omega_{k}\mid}}\; sinn\left\{\mid\Omega_{k}\mid^{1/2}
\times\int_{0}^{z}{dz\over\sqrt{(1+z)^{2}(1+\Omega_{m}z)-z(2+z)\Omega_{\Lambda}}}\right\}\;.
\end{equation}
In this equation, $\Omega_{k}$ represents the spatial curvature of the Universe---appearing
as a term proportional to the spatial curvature constant $k$ in the Friedmann equation.
In addition, $sinn$ is $\sinh$ when $\Omega_{k}>0$ and $\sin$  when $\Omega_{k}<0$.
For a flat Universe with $\Omega_{k}=0$, which is what we assume throughout this
paper, this equation simplifies to the form $R_{\rm h}/(1+z)$ times the integral.
The conversion from one cosmology to another therefore reduces predominantly to
an evaluation of Equations~(2.10) and (2.11).

But before we move on to the cluster samples, and carry out this comparison,
there is an additional ingredient one must incorporate into the calculation of
$f_{\rm gas}$, and this has to do with the measurement radius used to delimit
the volume over which $M_{\rm gas}$ and $M_{\rm tot}$ are determined.
This radius is selected by fixing the value of cluster overdensity,
\begin{equation}
\Delta\equiv {3M_{\rm tot}(<R_\Delta)\over 4\pi \rho_c(z_{\rm cluster})r_\Delta^3}\;,
\end{equation}
at its inferred redshift $z_{\rm cluster}$. Often, $\Delta$ is taken to be
2,500 (as in refs. \cite{1,2}; see next section),
but not always. (This is one of several reasons why we cannot combine
all of the available samples to carry out a single fitting procedure. As we shall see
in the next section, it is necessary to carry out the fitting for each individual
compilation of sources. Some discussion concerning which value is more trustworthy in
measuring $f_{\rm gas}$ appears in refs. \cite{30,31}.)
The third data set we are using \cite{3} assumes $\Delta=500$.

So there is an additional dependence of $f_{\rm gas}$ on the background
cosmology, beyond simply the factor appearing in Equation~(2.9), because
$r_{2,500}$ (or $r_{500}$) itself changes with the model. The reasoning
behind this is rather simple to understand \cite{2}. On the one hand,
we know that the total mass within $r_{2,500}$ is given by the expression
$M_{2,500}=(4\pi/3)r_{2,500}^3(2,500\,\rho_c)$. But since both $T_e$
and the density gradient in Equation~(2.7) are approximately constant in
the region of $\theta_{2,500}$, the hydrostatic equilibrium equation
gives $M_{2,500}\propto r_{2,500}$ (see Equation~2.8). These two
expressions should be equal, and since $\rho_c\sim H(z)^2$, we see
that $r_{2,500}\sim H(z)^{-1}$. Thus, the angle spanned by $r_{2,500}$
at $z$ is $\theta_{2,500}=r_{2,500}/d_A\sim (H\,d_A)^{-1}$. According
to ref. \cite{2}, the slope of $f_{\rm gas}(r/r_{2,500})$ in the
region of $r_{2,500}$ is $\eta\sim 0.214\pm0.022$ over their
sample range $0.7<r/r_{2,500}<1.2$, for the reference $\Lambda$CDM
model described in the introduction. Therefore, since the angle
subtended by $r_{2,500}$ changes with the cosmology, one
expects that $f_{\rm gas}\sim \theta_{2,500}^{-\eta}\sim
(H\,d_A)^\eta$, over and above the primary dependence given
in Equation~(2.9). This angular correction factor is close to unity
for all cosmologies and redshifts of interest, but ought to be included
for completeness.

Given the strong dependence of the inferred values of $M_{\rm gas}$
and $M_{\rm tot}$ on the assumed cosmology, the
data need to be recalibrated for each considered model.  However,
a procedure has been developed by the groups who analyze these
clusters, in which the data are reduced once for the reference
$\Lambda$CDM model, and then are fitted with modifications
to the reference model based on its differences with the cosmology
being tested. Specifically, the model fitted to the reference
$\Lambda$CDM data takes the form
\begin{equation}
f_{\rm gas}^{\rm model}=K\left[{H(z)\,d_A(z)\over [H(z)\,d_A(z)]^{\Lambda CDM}}\right]^\eta
\left[{d_A^{\Lambda CDM}(z)\over d_A(z)}\right]^{3/2}\;,
\end{equation}
where $K$ is a constant that includes a parametrization of the residual uncertainty in the
accuracy of the instrument calibration and X-ray modelling, and the factors in brackets
represent the two principal dependencies described above, i.e., on $d_A^{3/2}$
and $\theta_{2,500}^\eta$. In this expression, the variables with superscript
$\Lambda$CDM have values corresponding to the reference $\Lambda$CDM
model (see \S1 above), whereas the unlabeled parameters are those representing the new
cosmological model being tested (in this case, $R_{\rm h}=ct$). Sometimes, additional 
factors are added to this expression, e.g., representing the possible contribution 
from nonthermal pressure support, the $z$ dependence of the baryonic mass fraction 
in stars, and a possible evolutionary depletion of the baryon fraction measured at
$r_{2,500}$ as a consequence of the thermodynamic history of the gas \cite{2}. 
All these factors, however, appear to be very close to unity, and we will 
therefore not include them in our analysis.

\section{Theoretical Background}
The appropriate spacetime to use in any cosmological model is conveniently and
elegantly written in terms of the Friedmann-Robertson-Walker (FRW) metric, though
this does not tell us much about the cosmic equation of state (EOS), relating the total
energy density $\rho$ to its total pressure $p$. If the EOS were known,
the dynamical equations governing the Universal expansion could be solved exactly,
and the observations could then be interpreted unambiguously. Unfortunately, we
must rely on measurements and assumptions to pick $\rho$ and $p$. At the very minimum,
$\rho$ must contain matter $\rho_{\rm m}$ and radiation $\rho_{\rm r}$, which we see
directly, and an as yet poorly understand `dark' energy $\rho_{\rm de}$, whose presence
is required by a broad range of data including the Type Ia supernova Hubble
diagram \cite{32,33}.

However, as the measurements of the distance versus redshift continue to improve,
they appear to be creating more tension between theory and observations, rather than
providing us with a better indication of the dark-energy component, $p_{\rm de}=
w_{\rm de}\rho_{\rm de}$. For example, this is seen with the difficulty
$\Lambda$CDM has in accounting for the growth and evolution of high-$z$
quasars \cite{25} and high-$z$ galaxies \cite{27}.
It is also apparent from the $2.5\sigma$ disparity between the predictions
of $\Lambda$CDM and the precise measurements using the Alcock-Paczynski test
(${\cal D}(z) =d_A(z)H(z)/c$) applied to galaxy clusters \cite{34}.
Of the 3 measurements made to date, ${\cal D}(0.35)=0.286\pm0.025$,
${\cal D}(0.57)=0.436\pm0.052$, and ${\cal D}(2.34)=1.229\pm0.110$,
$\Lambda$CDM predicts ${\cal D}^{\Lambda{\rm CDM}}(0.35)=0.325$,
${\cal D}^{\Lambda{\rm CDM}}(0.57)=0.500$, and ${\cal D}^{\Lambda{\rm CDM}}(2.34)
=1.354$. By contrast, $R_{\rm h}=ct$ provides a much better accounting of these
data, with ${\cal D}^{R_{\rm h}=ct}(0.35)=0.300$, ${\cal D}^{R_{\rm h}=ct}(0.57)=
0.451$ and ${\cal D}^{R_{\rm h}=ct}(2.34)=1.206$. And as a third example,
the best-fit value of $H_0=67.3\pm1.2$ km s$^{-1}$ Mpc$^{-1}$ measured by
{\it Planck} \cite{35} is quite different from that ($\sim 70-72$ km
s$^{-1}$ Mpc$^{-1}$) inferred from low-redshift measurements, e.g., using
the Type Ia SN Hubble diagram.

$\Lambda$CDM assumes that dark energy is a cosmological constant $\Lambda$ with
$w_{\rm de}\equiv w_\Lambda=-1$, and therefore $w=(\rho_{\rm r}/3-\rho_\Lambda)/\rho$.
This model does quite well explaining many of the observations, but such a scenario
is inadequate to explain all of the nuances seen in cosmic evolution and the
growth of structure. For example, insofar as the CMB fluctuations measured with
both WMAP \cite{36} and {\it Planck} \cite{35} are concerned,
there appears to be unresolvable tension between the predicted and measured angular
correlation function \cite{26,37,38,39}.
Also, the observed galaxy distribution function appears to be scale-free,
whereas the matter distribution expected in $\Lambda$CDM has a different form on different
spatial scales. The fine tuning required to resolve this difference led Watson et al. \cite{40}
to characterize the galactic matter distribution function as a `cosmic coincidence.'
(We note, however, that the galaxy correlation function may be a poor indicator
of the matter distribution itself, since the former depends on the still uncertain nature
of galaxy formation, in addition to the underlying cosmology.)  Such difficulties
are compounded by $\Lambda$CDM's predicted redshift-age relation,
which does not appear to be consistent with the growth of quasars
at high redshift \cite{25}, nor the very early appearance of galaxies at
$z\gtrsim 10$ \cite{27}.

The $R_{\rm h}=ct$ Universe is another FRW cosmology that has much in common
with $\Lambda$CDM, but includes an additional ingredient motivated by several
theoretical and observational arguments \cite{15,16,17}. Like $\Lambda$CDM, 
it also adopts the equation of state
$p=w\rho$, with $p=p_{\rm m}+ p_{\rm r}+p_{\rm de}$ and $\rho=\rho_{\rm m}+
\rho_{\rm r}+\rho_{\rm de}$, but goes one step further by specifying that
$w=(\rho_{\rm r}/3+ w_{\rm de}\rho_{\rm de})/\rho=-1/3$ at all times. One might
come away with the impression that these two prescriptions for the equation of
state cannot be consistent. But in fact nature is telling us that if we ignore the
constraint $w=-1/3$ and instead proceed to optimize the parameters in $\Lambda$CDM by
fitting the data, the resultant value of $w$ averaged over a Hubble time
is actually $-1/3$ within the measurement errors \cite{15,17}.
In other words, though $w=(\rho_{\rm r}/3-\rho_\Lambda)/\rho$
in $\Lambda$CDM cannot be equal to $-1/3$ from one moment to the next, its value averaged
over the age of the Universe is equal to what it would have been in $R_{\rm h}=ct$.

This result does not necessarily prove that $\Lambda$CDM is an incomplete version of
$R_{\rm h}=ct$, but it does seem to suggest that the inclusion of the additional
constraint $w=-1/3$ might render its predictions closer to the data. In $R_{\rm h }=ct$,
this condition on the total equation of state is required in order to maintain a
constant expansion rate $a(t)\propto t$. Thus, the principal difference between
$\Lambda$CDM and $R_{\rm h}=ct$ is that, whereas one must first assume the constituents and
their equations of state in $\Lambda$CDM and then infer its expansion rate, the
Universe's dynamics in $R_{\rm h}=ct$ is completely specified before one begins
to speculate on its contents.

Nonetheless, both $\Lambda$CDM and $R_{\rm h}=ct$ face similar limitations
when it comes to the essential ingredients in the cosmic fluid, such as the
nature of dark matter or dark energy. In $\Lambda$CDM, one of the most important
assumptions is that $\rho_b/\rho_{d}$ is constant with redshift (and therefore
time). Of course, the distribution of halos, and eventually galaxies
and clusters, depends on the background expansion rate (for a recent
set of simulations, see ref. \cite{14}). However, the many calculations
carried out to date suggest that the gas mass fraction in clusters is insensitive
to the underlying cosmology. 

There is actually a good, sound reason for the relative insensitivity 
of the structure and content of a condensing halo to the choice of underlying 
cosmology. One may understand this basic outcome in the context of Birkhoff's 
theorem and its corollary \cite{15,19}, according to which the spacetime inside 
of a spherical shell in an otherwise isotropic distribution of mass and energy 
is completely independent of the exterior region. Because of spherical symmetry, 
all contributions to the spacetime curvature within this shell cancel completely. 
Thus, once an overdense perturbation in the background density begins to be 
self-gravitating and forms a bound system, its subsequent evolution proceeds 
under its own gravity, independently of the surrounding medium, even in an 
infinite universe. The expansion rate exterior to the contracting halo 
therefore has little influence on the eventual structure and content of 
the collapsing region. All the simulations confirm this basic result---actually 
going even farther and showing that the inclusion of additional 
astrophysical effects, such as radiative cooling, have only a minimal 
impact on the results.

In other words, because of Birkhoff's theorem, if $f_{\rm gas}$ is more 
or less constant in any one model, it is expected to be similarly 
constant in all cosmological models, but we would measure it to be
independent of redshift only if the correct geometry were assumed in
the data analysis. This is what makes it such a potentially powerful 
probe of the cosmology. Insofar as the $R_{\rm h}=ct$ Universe is concerned,
detailed hydrodynamical simulations of structure formation do not 
yet exist. But because of Birkhoff's theorem and the insensitivity of
the halo evolution to the external expansion rate, we can already
start to examine the possibility that $f_{\rm gas}$ may be constant
in this cosmology as well, with more in-depth analysis to follow after
comprehensive structure formation simulations will have been completed. 

Nonetheless, to make a model comparison viable, we need to consider
several essential constraints. At the very minimum, the $R_{\rm h}=ct$ 
Universe must also contain baryonic matter, radiation, dark matter, and 
some form of dark energy, though we already know that this could not 
be a cosmological constant. What we do know, however, is that no matter 
what these ingredients turn out to be, they must always partition 
themselves in such a way as to maintain the total equation of state 
$p=-\rho/3$. The idea that the internal chemisty of a system adjusts 
to macroscopic constraints is not uncommon. For example, we already 
have such a situation with $\Lambda$CDM, where the partitioning of 
baryonic matter and radiation in the early universe follows the prescribed
redshift evolution in temperature $T(z)$. Moreover, since the existence
of dark matter and dark energy presumably implies physics beyond the
standard model, it's quite possible that this early partitioning
of the constituents in $\Lambda$CDM involves other components, in
addition to baryonic matter and radiation.

Recently, the validity of the $R_{\rm h}=ct$ model was questioned
on the basis that its equation of state could not be consistent with
the conservation of matter during the Universe's expansion \cite{42}.
This argument took the opposite approach to what we have just
described, i.e., it abandoned the equation of state $p=-\rho/3$
and instead replaced it with a $\Lambda$CDM-like sum of the
equations of state of individual constituents, but with the
added proviso that matter could not be created or annihilated
once it appeared on the scene. This begs the question of how
matter could have been created in the first place, not to mention
how such an unmotivated constraint could be consistent with what
we believe happened with $\Lambda$CDM in the early Universe, when
matter and radiation (and possibly other as yet unknown fields)
transformed back and forth into each other prior to, and subsequent
to, the period of inflation. It's always interesting to explore 
the viability of such variants to the basic model, but one should 
not interpret their results as being meaningful to the 
$R_{\rm h}=ct$ Universe, which does not incorporate the 
conservation of matter as one of its basic ingredients. 
The only condition essential to this cosmology is the 
total equation of state $p=-\rho/3$, which no doubt will
impact how we interpret physics beyond the standard model.

In our recent analysis of the Epoch of Reionization \cite{41}, 
we considered the possibility that the dark matter and baryonic 
densities might have evolved separately of each other at high 
redshifts, i.e., $6\lesssim z\lesssim 15$. However, it is not
clear whether such a trend might continue to lower redshifts.
There is clearly much to learn from future developments in
particle physics, for both $\Lambda$CDM and $R_{\rm h}=ct$. But
insofar as understanding the redshift dependence of the gas mass
fraction $f_{\rm gas}$ is concerned, we will here make the
simplest minimal assumption for both $R_{\rm h}=ct$ and $\Lambda$CDM, 
which is that the baryonic fraction $\rho_b/\rho_{d}$ remains 
approximately constant for $z\lesssim 2$. The results of this paper 
are contingent upon the validity of this assumption.

\section{The Principal Data Sets}
In order to compare the predictions of the $R_{\rm h}=ct$ Universe against
those of the reference $\Lambda$CDM model, we consider three samples
of galaxy clusters whose gas mass fractions have been measured using
X-ray surface brightness observations. The LaRoque et al. sample \cite{1}
consists of 38 massive clusters lying in the redshift range $0.14<z<0.89$,
and were obtained from {\it Chandra} X-ray and OVRO/BIMA interferometric
Sunyaev-Zeldovich Effect measurements. In order to study the
dependence of their analysis on the assumed model for the cluster gas
distribution, taking into account the possible presence of a cooling flow,
these authors considered three different models for the plasma
profile: (1) an isothermal $\beta$-model fit jointly to the X-ray data
at radii beyond 100 kpc and to all of the SZE data, (2) a nonisothermal
double $\beta$-model in hydrostatic equilibrium, fit jointly to all of the
X-ray and SZE data, and (3) an isothermal $\beta$-model fit only to
the SZE spatial data. In this paper, we consider the results of models
(1) and (2) only, since the number of clusters appropriate for the third
case was noticeably smaller than the others. The single isothermal
$\beta$-model with the central 100 kpc excised seemed to work
quite well, since the cut was large enough to exclude the
cooling region in cool-core clusters while keeping a sufficient
number of photons to enable the mass modelling. The  more
sophisticated double-$\beta$ model was designed to take into
account (non-isothermal) temperature profiles, and was
developed to assess the biases arising from the isothermal
assumption and the effects of core exclusion in the first
model. As we will see in the next section, the fits suggest
that both of these approaches work quite well, and provide
mutually consistent results.

Our second sample is taken from Allen et al. \cite{2}, who compiled a catalog
of 42 hot ($kT_e>5$ keV), X-ray luminous, dynamically relaxed galaxy
clusters spanning the redshift range $0.05<z<1.1$. Their measurements
were also based on {\it Chandra} observations, and like ref. \cite{1},
these authors also adopted a canonical measurement radius of $r_{2,500}$,
whose value they determined directly from the {\it Chandra} data. Ten
of these clusters are in common with a subset of the LaRoque sample,
and there is good agreement for the best-fitting results in this sub-sample
at $r_{2,500}$ between the two groups, though the mass fractions
measured by LaRoque et al. are on average about $6$ percent
higher than those reported in ref. \cite{1} for the systems
in common.

The Ettori et al. \cite{3} sample is the biggest of the three, containing
52 X-ray luminous galaxy clusters (also observed with {\it Chandra})
in the redshift range $0.3<z<1.273$, merged with 8 additional objects
at $0.06<z<0.23$, with a gas temperature $>4$ keV \cite{30}.
Note, however, that
although the reference cosmology (as described in \S1 above)
is identical for all three samples, ref. \cite{3} decided to use
an overdensity of 500 (instead of 2,500) to define the outer radius of
their mass determination region. Strictly speaking, this means that
$\eta$ (see Equation~2.13) could be different from the value ($\sim 0.2$)
applied to the other two samples, but for the sake of simplicity, we
will use the same value throughout our analysis. Since this index
is presumably much smaller than one, the impact of
this approximation on our results is expected to be smaller than,
e.g., the errors on the sample means of the $f_{\rm gas}$ values.
Another technical difference among the samples is that both the
Allen et al. \cite{2} and Ettori et al. \cite{3} measurements of the
mass fraction $f_{\rm gas}$ are based strictly on the assumption
of isothermality and hydrostatic equilibrium, in constrast to 
ref. \cite{1}, which considered both isothermal and nonisothermal 
models.  Such differences in the handling of the various samples
precludes any possibility of merging them into a single, bigger sample,
thereby improving the statistics. On the other hand, the fact that
the approaches were somewhat different lends some credence to the
results when they agree with each other within the errors.

\section{A Direct Comparison between $\Lambda$CDM and $R_{\rm h}=ct$}
Non-radiative simulations of large clusters suggest that $f_{\rm gas}$ should
be approximately constant with redshift \cite{4,5,14}, at least for $z\lesssim 2$.
Let us now compare the measured values of $f_{\rm gas}$, based on the
reference $\Lambda$CDM model, with those re-calibrated for the
$R_{\rm h}=ct$ Universe using Equation~(2.13). For each sample, we calculate
the $\chi^2$ function,
\begin{equation}
\chi^2=\sum_{i=1}^N{\left(f_{{\rm gas},i}-f_{\rm gas}\right)^2
\over \sigma_i^2+\sigma_{f}^2}\;,
\end{equation}
where $N$ is the sample size, $f_{{\rm gas},i}$ and $\sigma_i$ are the single
gas mass fraction measurements and their relative errors, $f_{\rm gas}$
is the constant gas fraction to be optimized while finding the best fit to
the $f_{{\rm gas},i}$ values, and $\sigma_{f}$ is its error, calculated
from the population standard deviation \cite{3}.

\begin{figure}[!h]
\centering\includegraphics[width=0.6\linewidth]{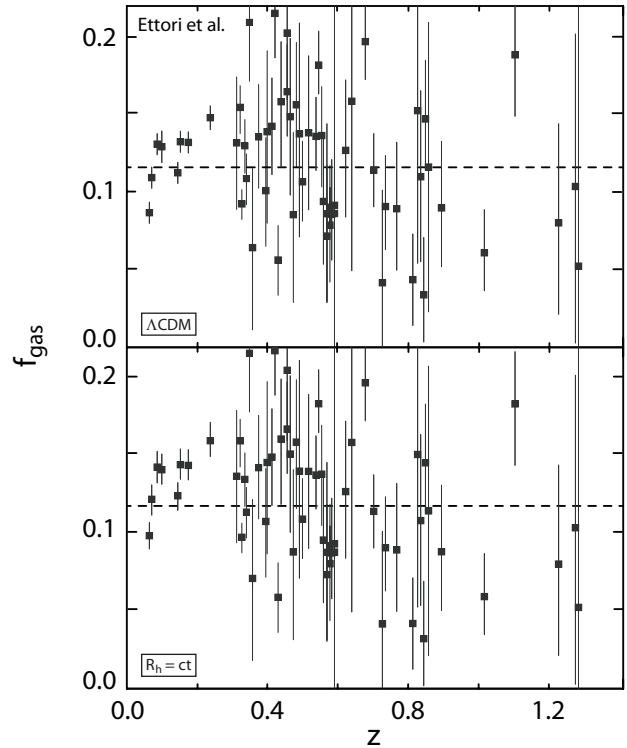}
\caption{The apparent variation of the X-ray gas mass fraction measured within
$r_{500}$ as a function of redshift for the 60 clusters in Ettori et al. (2009). In the 
upper panel are the values for the reference $\Lambda$CDM model ($\Omega_{\rm m}=0.3$,
$h\equiv H_0/100\;{\rm km}\;{\rm s}^{-1}\;{\rm Mpc}^{-1}=0.7$, and $w_{\rm de}=-1$).
In the lower panel, are the values for the $R_{\rm h}=ct$ Universe, using the same Hubble
constant $h$. Both results are consistent with the expectation of a
constant $f_{\rm gas}(z)$ (dashed lines) from simulations. The reference
$\Lambda$CDM fit yields $f_{\rm gas}=0.114\pm0.041$, with $\chi^2_{\rm dof}=0.57$
for $58$ degrees of freedom, while fitting with the $R_{\rm h}=ct$ Universe yields
$f_{\rm gas}=0.117\pm0.043$, with $\chi^2_{\rm dof}=0.56$ for $60$
degrees of freedom. By comparison, the cosmic ratio $\rho_b/(\rho_b+\rho_{d})$ measured
by {\it Planck} is $0.155\pm0.006$ \cite{35}, and $0.166\pm0.013$ measured
by WMAP-9 \cite{39}.}
\label{figure1}
\end{figure}

The Hubble constant itself does not affect the comparison between the two models.
Therefore, fits to the data using the $R_{\rm h}=ct$ Universe have no free parameters.
One can see this directly from Equation~(2.10), in which the removal of $R_{\rm h}$, i.e.,
the Hubble constant, leaves no flexibility at all for the angular distance as a function of
redshift. $\Lambda$CDM, on the other hand, has anywhere from 2 to 6 free parameters,
in addition to $H_0$, depending on how one chooses to treat the dark energy and its
equation of state. Here, we conservatively take the minimum number, i.e., 2, these
being the value of $\Omega_m$ and $w_\Lambda$, the two parameters (besides
$H_0$) used to calculate $f_{\rm gas}$ in the three samples.

To facilitate a quick visual comparison between the various models, we show in
figures~1-3 the data obtained for the reference $\Lambda$CDM cosmology,
paired with the same set of data re-calibrated for the $R_{\rm h}=ct$ Universe. As
described above, this re-calibration is not necessary to produce the fits and
their $\chi^2$ values, and is carried out solely for the purpose of yielding an
immediate visual impact of the differences between the two.

\begin{figure}[!h]
\centering\includegraphics[width=0.8\linewidth]{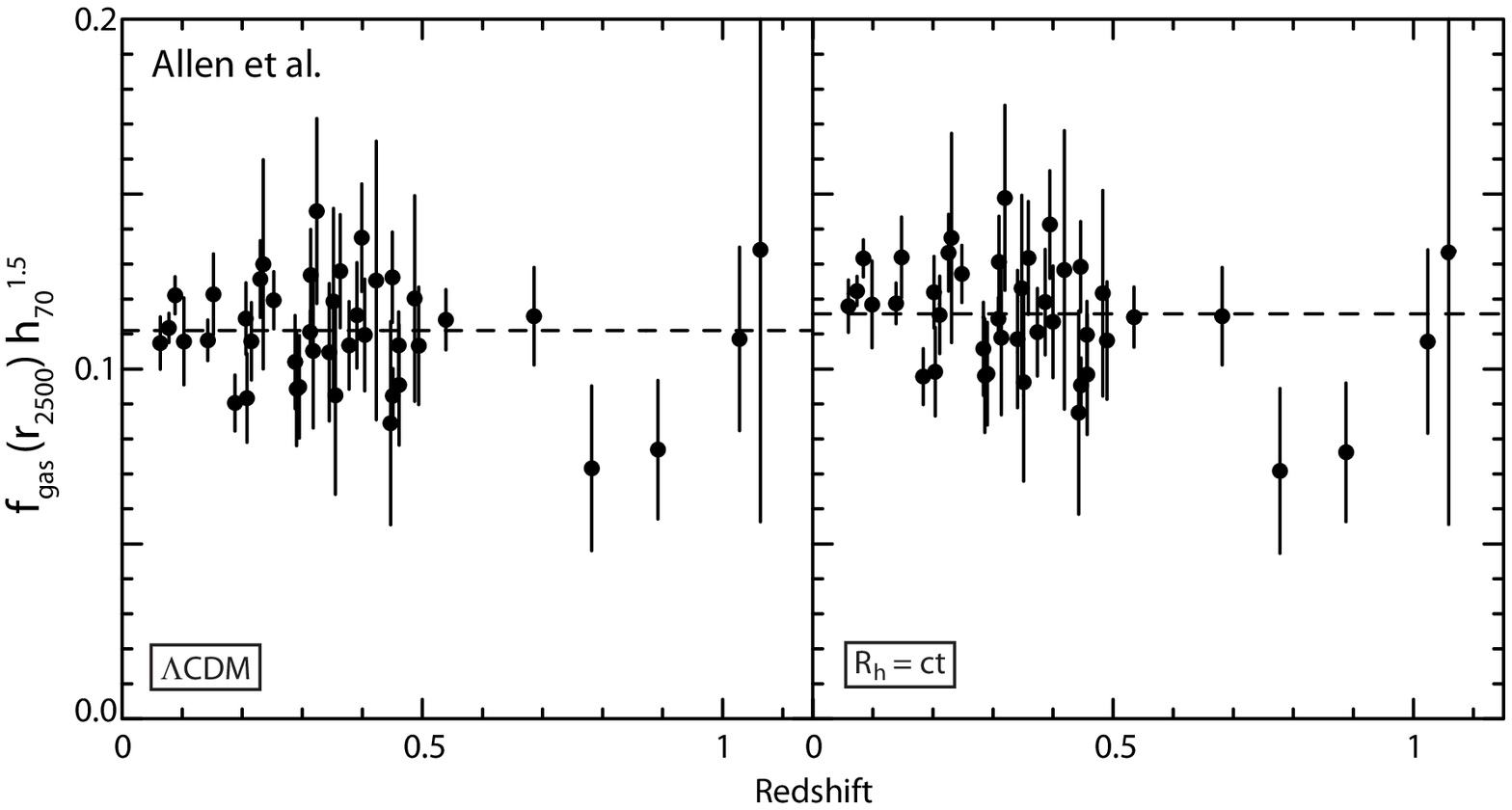}
\caption{The apparent variation of the X-ray gas mass fraction measured within
$r_{2500}$ as a function of redshift for the 42 clusters in Allen et al. \cite{2}. On
the left are the values for the reference $\Lambda$CDM model ($\Omega_{\rm m}=0.3$,
$h\equiv H_0/100\;{\rm km}\;{\rm s}^{-1}\;{\rm Mpc}^{-1}=0.7$, and $w_{\rm de}=-1$).
On the right, are the values for the $R_{\rm h}=ct$ Universe, using the same Hubble constant $h$
(though $h$ has no effect on this plot). Both results are consistent with the expectation of a
constant $f_{\rm gas}(z)$ (dashed lines) from simulations. The reference
$\Lambda$CDM fit yields $f_{\rm gas}=0.110\pm0.016$, with $\chi^2_{\rm dof}=0.42$
for $40$ degrees of freedom, while fitting with the $R_{\rm h}=ct$ Universe yields
$f_{\rm gas}=0.116\pm0.017$, with $\chi^2_{\rm dof}=0.45$ for $42$
degrees of freedom. By comparison, the cosmic ratio $\rho_b/(\rho_b+\rho_{d})$ measured
by {\it Planck} is $0.155\pm0.006$ \cite{35}, and $0.166\pm0.013$ measured
by WMAP-9 \cite{39}.}
\end{figure}

In all three samples, both cosmologies are consistent with the expectation of a constant
$f_{\rm gas}$, though slight differences emerge for the values of $f_{\rm gas}$
and $\chi^2_{\rm dof}$. However, the fact that the number of free parameters is
different between these models leads to significantly different likelihoods of either
being closer to the correct cosmology, as we shall describe in the next section.
But for now, starting with the most recently measured sample \cite{3},
we find that the reference $\Lambda$CDM model yields $f_{\rm gas}=0.114\pm0.041$,
with $\chi^2_{\rm dof}=0.57$, for $60-2=58$ degrees of freedom. By comparison,
the fit using $R_{\rm h}=ct$ results in $f_{\rm gas}=0.117\pm0.043$, with
$\chi^2_{\rm dof}=0.56$, for $60$ degrees of freedom. The fits are comparable,
though with very slight differences in the average gas mass fraction (see figure~1).

For the Allen et al. \cite{2} sample (figure~2), we find using this approach that the
reference $\Lambda$CDM model yields $f_{\rm gas}=0.110\pm0.016$
with $\chi^2_{\rm dof}=0.42$ for $42-2=40$ degrees of freedom, while
fitting with $R_{\rm h}=ct$ gives $f_{\rm gas}=0.116\pm0.017$
and $\chi^2_{\rm dof}=0.45$ for 42 degrees of freedom.

\begin{figure}[!h]
\centering\includegraphics[width=0.6\linewidth]{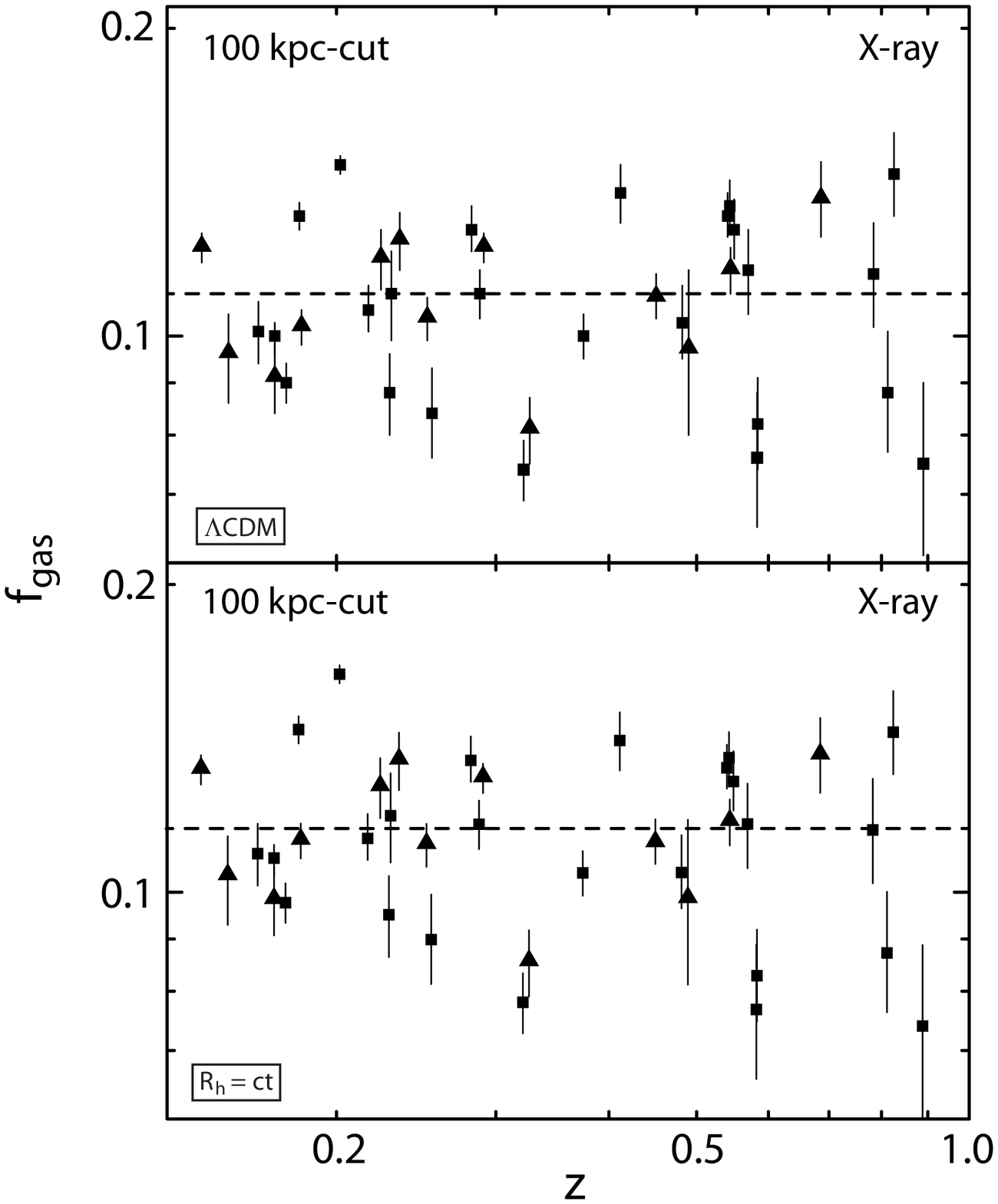}
\caption{\hskip-0.05in{\bf a}. The apparent variation of the X-ray gas mass fraction measured within
$r_{2500}$ as a function of redshift for the 38 clusters in ref. \cite{1}, based
on {\it Chandra} X-ray data. Triangles: the cool-core subsample; Squares: the non-cool-core
subsample (see text). Both cases assume a single isothermal $\beta$-model with
the central 100 kpc excised. {\it Upper panel:} the values are for the reference $\Lambda$CDM model
($\Omega_{\rm m}=0.3$, $h\equiv H_0/100\;{\rm km}\;{\rm s}^{-1}\;{\rm Mpc}^{-1}=0.7$,
and $w_{\rm de}=-1$). {\it Lower panel:} The $R_{\rm h}=ct$ Universe. Both results are consistent
with the expectation of a constant $f_{\rm gas}(z)$ (dashed lines) from simulations.
The reference $\Lambda$CDM fit yields $f_{\rm gas}=0.108\pm0.020$, with $\chi^2_{\rm dof}=0.88$
for $36$ degrees of freedom (upper panel), and the $R_{\rm h}=ct$ fit yields $f_{\rm gas}=
0.114\pm0.021$, with $\chi^2_{\rm dof}=0.84$ for $38$ degrees of freedom (lower panel).
By comparison, the cosmic ratio $\rho_b/(\rho_b+\rho_{d})$ measured
by {\it Planck} is $0.155\pm0.006$ \cite{35}, and $0.166\pm0.013$ measured
by WMAP-9 \cite{39}.}
\end{figure}

\setcounter{figure}{2}
\begin{figure}[!h]
\centering\includegraphics[width=0.6\linewidth]{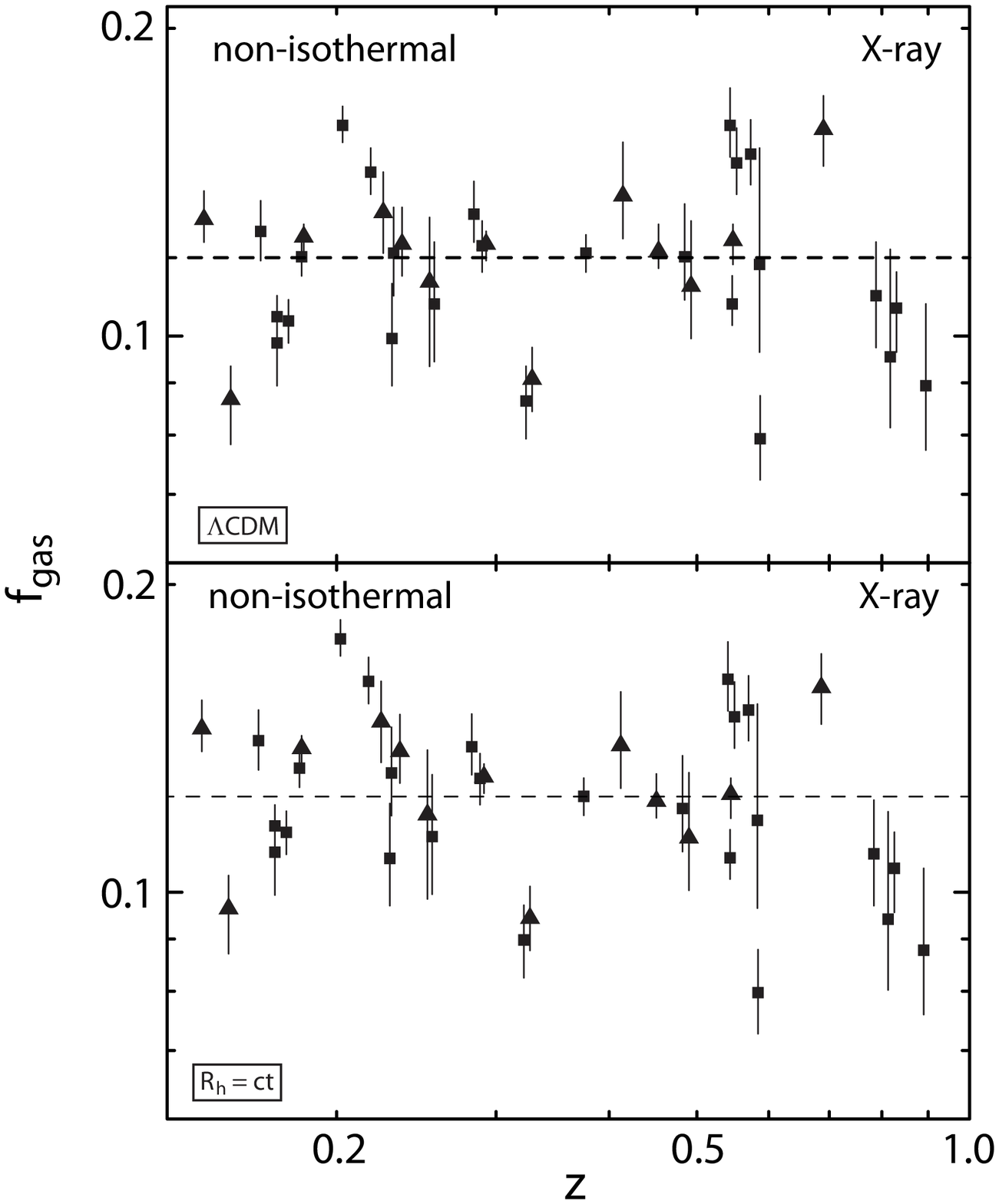}
\caption{\hskip-0.05in{\bf b}. Same as Fig.~3.a, except now assuming a nonisothermal $\beta$-model.
Triangles and squares have the same meaning.
{\it Upper panel:} the values are for the reference $\Lambda$CDM model.
{\it Lower panel:} the $R_{\rm h}=ct$ Universe. The fits yield $f_{\rm gas}=0.118\pm0.021$,
with $\chi^2_{\rm dof}=0.82$ for $36$ degrees of freedom (upper panel), and
$f_{\rm gas}=0.124\pm0.022$, with $\chi^2_{\rm dof}=0.86$ for $38$ degrees
of freedom (lower panel). By comparison, the cosmic ratio $\rho_b/(\rho_b+\rho_{d})$ measured
by {\it Planck} is $0.155\pm0.006$ \cite{35}, and $0.166\pm0.013$ measured
by WMAP-9 \cite{39}.}
\end{figure}

In the LaRoque et al. \cite{1} sample, we consider the isothermal (with a
100-kpc cut) cases (figures~3.a and 3.c) separately from the non-isothermal
cases (figures~3.b and 3.d), and also the X-ray observed gas mass fractions
(figures~3.a and 3.b) separately from those obtained via measurements of the
SZE (figures~3.c and 3.d). In these figures, the data include both cool-core
(triangles) and non-cool-core subsamples (squares). This sample includes a
subgroup of clusters with bright and sharply peaked cores; they are referred
to as cool-core clusters because the sharply peaked X-ray emission is indicative
of strong radiative cooling in cluster core. The individual values of $f_{\rm gas}$
and $\chi^2_{\rm dof}$ are quoted in the figure captions, and range over
$f_{\rm gas}\sim 0.108\pm0.020$ to $0.120\pm0.032$ with $\chi^2_{\rm dof}\sim
0.59-0.88$ for the reference $\Lambda$CDM model, and $f_{\rm gas}
\sim 0.114\pm0.021$ to $0.124\pm0.022$ with $\chi^2_{\rm dof}\sim
0.57-0.86$ for the $R_{\rm h}=ct$ Universe.

\section{Discussion and Conclusions}
The results presented in the previous section demonstrate that both
the reference $\Lambda$CDM cosmology and the $R_{\rm h}=ct$
Universe are consistent with a constant value of the cluster gas mass
fraction with increasing redshift, under the underlying assumption
that the baryonic fraction $\rho_b/\rho_d$ has remained constant
during the most significant period of structure formation (i.e.,
$z\lesssim 3-4$).

The fact that the $R_{\rm h}=ct$ Universe fits these data so well
is probably the reason why some previous work with clusters had already
hinted at a possible deviation from accelerated expansion, even though
$R_{\rm h}=ct$ was not known or used in those studies \cite{43}.
Using the 42 measurements from ref. \cite{2}, these authors concluded
that cosmic acceleration in the context of $\Lambda$CDM could have already
peaked and that we might be witnessing a slowing down. This effect was
also found previously by ref. \cite{44} using supernova data.

But the process of selecting the most likely correct model also takes into
account the number of free parameters. The likelikhood of either
$R_{\rm h}=ct$ or $\Lambda$CDM being closer to the ``true" model may
be determined from the model selection criteria discussed extensively in
ref. \cite{23}. A commonly used criterion
in cosmology is the Akaike Information Criterion (AIC) \cite{45,46,47},
which prefers models with few parameters to those
with many, unless the latter provide a substantially better fit to the data.
This avoids the possibility that by using a greater number of parameters, one
may simply be fitting the noise.

The AIC is given by ${\rm AIC} = \chi^2 + 2\,k$,
where $k$ is the number of free parameters. Among two models $\mathcal{M}_1$ and
$\mathcal{M}_2$ fitted to the data, the one with the least resulting AIC is assessed as the
one more likely to be ``true.''  If ${\rm AIC}_i$ comes from model~$\mathcal{M}_i$, the
unnormalized confidence that $\mathcal{M}_i$~is true is the ``Akaike
weight'' $\exp(-{\rm AIC}_i/2)$.  Informally, $\mathcal{M}_i$~has
likelihood
\begin{equation}
\label{eq:lastAIC}
{\cal L}(\mathcal{M}_i)=
\frac{\exp(-{\rm AIC}_i/2)}
{\exp(-{\rm AIC}_1/2)+\exp(-{\rm AIC}_2/2)}
\end{equation}
of being closer to the correct model.
A lesser known alternative, though based on similar arguments, is the
Kullback Information Criterion (KIC), which takes into account the fact that the PDF's
of the various competing models may not be symmetric. The unbiased estimator
for the symmetrized version \cite{48} is given by ${\rm KIC} = \chi^2 + 3\,k$,
very similar to the AIC, but clearly strengthening the dependence on the
number of free parameters (from $2k$ to $3k$).
The Bayes Information Criterion (BIC) is perhaps the best known of the three,
and represents an asymptotic ($N\to\infty$) approximation to the outcome of a
conventional Bayesian inference procedure for deciding between models
\cite{49}. This criterion is defined by ${\rm BIC} = \chi^2 + (\ln N)\,k$,
and clearly suppresses overfitting very strongly if $N$~is large.

\setcounter{figure}{2}
\begin{figure}[!h]
\centering\includegraphics[width=0.6\linewidth]{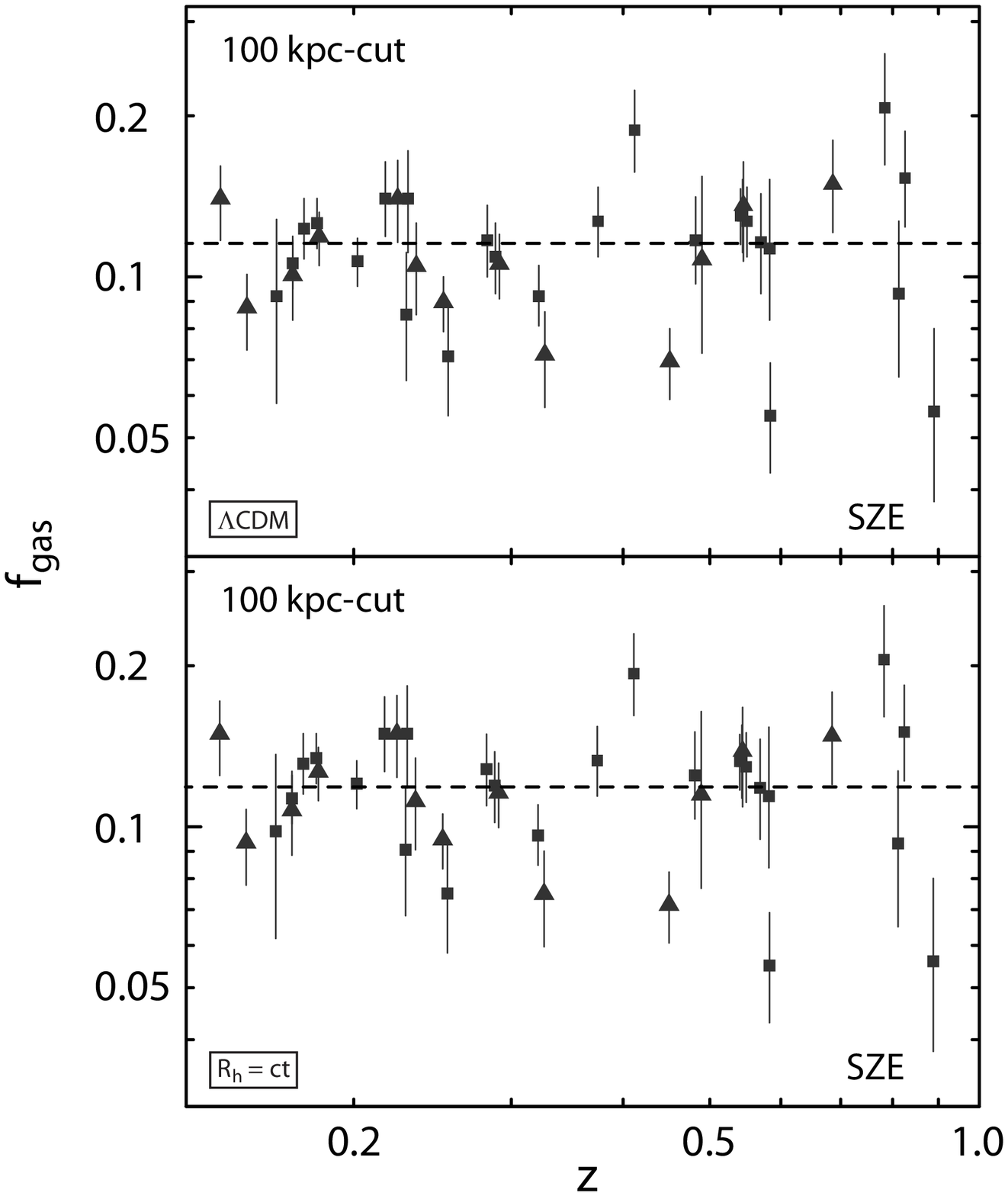}
\caption{\hskip-0.07in{\bf c}. Same as Fig.~3.a, except here for the SZE data.
Triangles and squares have the same meaning.
The reference $\Lambda$CDM fit yields $f_{\rm gas}=0.114\pm0.031$, with $\chi^2_{\rm dof}=0.59$
for $36$ degrees of freedom (upper panel), and the $R_{\rm h}=ct$ fit yields
$f_{\rm gas}=0.117\pm0.032$, with $\chi^2_{\rm dof}=0.57$ for $38$
degrees of freedom (lower panel). By comparison, the cosmic ratio $\rho_b/(\rho_b+\rho_{d})$ measured
by {\it Planck} is $0.155\pm0.006$ \cite{35}, and $0.166\pm0.013$ measured
by WMAP-9 \cite{39}.}
\end{figure}

\setcounter{figure}{2}
\begin{figure}[!h]
\centering\includegraphics[width=0.6\linewidth]{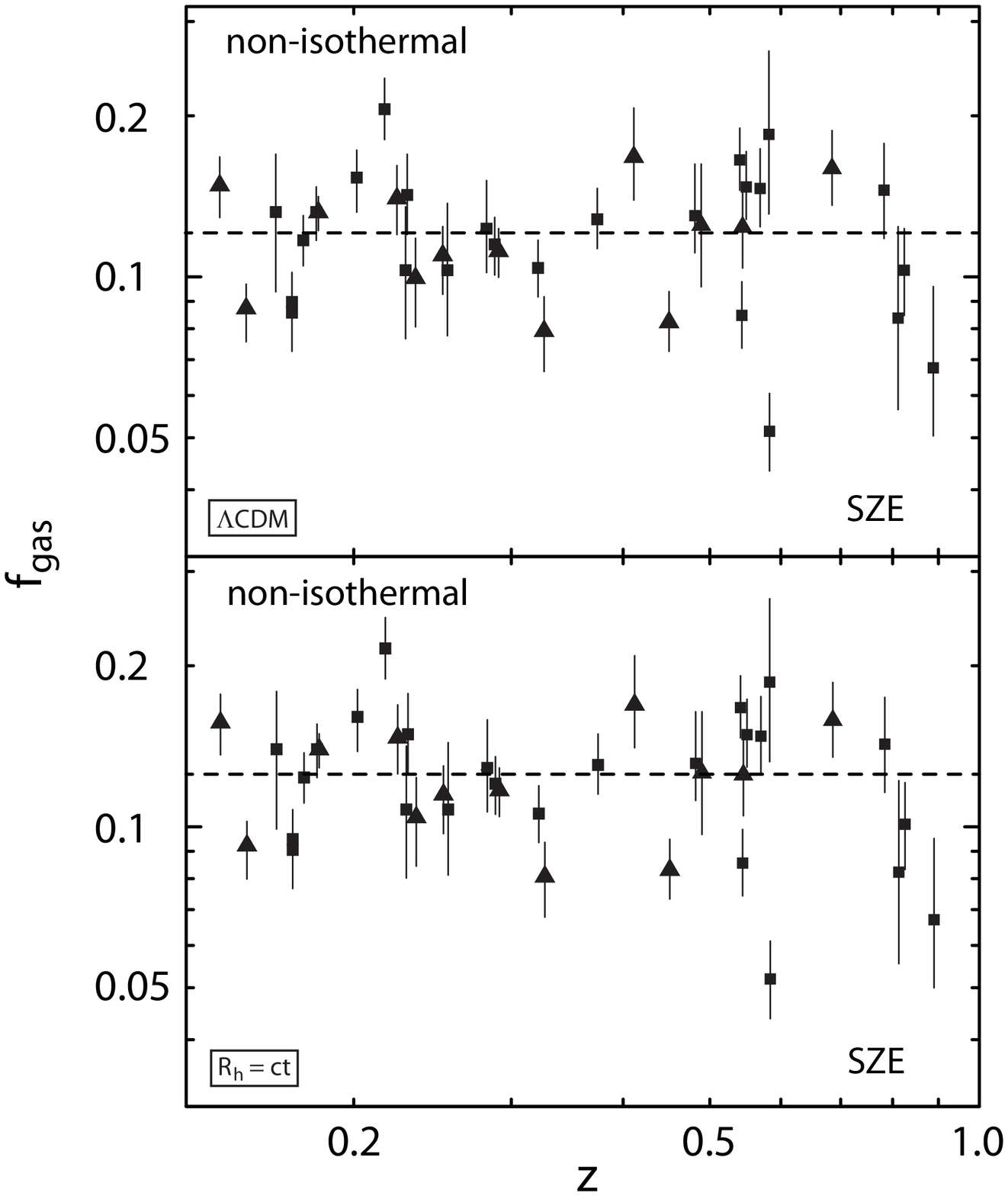}
\caption{\hskip-0.07in{\bf d}. Same as Fig.~3.b, except here for the SZE data.
Triangles and squares have the same meaning. The $\Lambda$CDM
fit yields $f_{\rm gas}=0.120\pm0.032$, with $\chi^2_{\rm dof}=0.67$
for $36$ degrees of freedom (upper panel), and $R_{\rm h}=ct$ yields
$f_{\rm gas}=0.123\pm0.033$, with $\chi^2_{\rm dof}=0.67$ for $38$ 
degrees of freedom (lower panel). By comparison, the cosmic ratio 
$\rho_b/(\rho_b+\rho_{d})$ measured by {\it Planck} is $0.155\pm0.006$ 
\cite{35}, and $0.166\pm0.013$ measured by WMAP-9 \cite{39}.}
\end{figure}

For the fits discussed in the previous section, these three model selection
criteria result in the likelihoods shown in Table 1. The point of
listing all three criteria is not so much to dwell on which of these may or
may not reflect the importance of free parameters but, rather, to demonstrate
that there is general agreement among them---the most commonly used
model-selection tools in the literature---that the cluster gas mass fraction
data favor the $R_{\rm h}=ct$ Universe over $\Lambda$CDM.
In the case of BIC, considered to be the
most reliable among them \cite{45,46}, the
difference in likelihoods is overwhelming ($\sim 95\%$ to $\sim 5\%$).
This effect is considered to be `strong' when using these
criteria. Note also that, in spite of the fact that the Ettori et al.
\cite{3} sample uses a different over-density ratio $\Delta$ than those
of refs. \cite{1,2}, the likelihood
comparisons in Table 1 are all quite similar and consistent with each other.
This may be a fortuitous result because simulations have shown that the baryon
fraction in clusters can vary depending on the level of concentration towards
the core, as a result of several astrophysical effects, including radiative
cooling and feedback from winds (see also ref. \cite{50}).
It appears that the level of measurement
precision we have currently is not sufficient to discern between these
two values of $\Delta$. However, the fact that theory predicts some change
in $f_{\rm gas}$ with spatial scale suggests that future observations may
need to be interpreted more carefully when the baryon fraction is used to
do cosmological model comparisons.

\begin{table*}[h]
\center
\centerline{Table 1. Likelihood Estimation}
  \begin{tabular}{lccc}
    \hline
{Data Source} & {Criterion} & {$\Lambda$CDM} & {$R_{\rm h}=ct$}\\
\hline
Ref.~\cite{3}&AIC&$18\%$&$82\%$\\
&KIC&$8\%$&$92\%$\\
&BIC&$3\%$&$97\%$\\
Ref.~\cite{2}&AIC&$34\%$&$64\%$\\
&KIC&$11\%$&$89\%$\\
&BIC&$6\%$&$94\%$\\
Ref.~\cite{1}&&\\
X-ray 100 kpc cut&AIC&$13\%$&$87\%$\\
&KIC&$6\%$&$94\%$\\
&BIC&$3\%$&$97\%$\\
X-ray nonthermal&AIC&$66\%$&$34\%$\\
&KIC&$19\%$&$81\%$\\
&BIC&$13\%$&$87\%$\\
SZE 100 kpc cut&AIC&$11\%$&$89\%$\\
&KIC&$5\%$&$95\%$\\
&BIC&$3\%$&$97\%$\\
SZE nonthermal&AIC&$21\%$&$79\%$\\
&KIC&$9\%$&$91\%$\\
&BIC&$5\%$&$95\%$\\
\hline
  \end{tabular}
\end{table*}

Interestingly, these likelihoods are similar to those inferred from our
analysis of the cosmic chronometer data \cite{23}, and
from our consideration of the gamma-ray burst Hubble Diagram \cite{24}.
Together, these tests are beginning to paint a consistent
picture. At best, $\Lambda$CDM may do as well as $R_{\rm h}=ct$
in accounting for some of the data, though at a cost---the need to
include a larger number of free parameters. But in some cases,
such as the cosmic chronometers, the $\chi^2$ of the $\Lambda$CDM
fit is inferior to that of $R_{\rm h }=ct$, even though the former
has a larger number of unrestricted variables.

Insofar as the use of cluster gas mass fractions to probe the cosmological
expansion is concerned, there is considerable room for improvement beyond
the current situation. For example, as the precision of the observations
continues to improve, and as the hydrodynamic simulations gain in sophistication 
and complexity, it is becoming more apparent that the adoption of a purely 
constant fraction $f_{\rm gas}$ may be an over-simplification. This ratio 
apparently changes with radius in any given cluster and, worse, may not be 
uniformly constant (as evidenced in part by the observed scatter) across 
a chosen sample (see, e.g., refs~\cite{52,53}). At the very minimum,
these effects call into question the use of different $\Delta$'s to 
calculate $f_{\rm gas}$. Part of the difficulty is that the best 
spatially-resolved data are not fully consistent with the assumption 
of hydrostatic equilibrium, which was used to infer the mass 
of the X-ray emitting plasma. Recently, substantial progress
has been made with cluster observations, driven by weak gravitational
lensing, which does not depend on the dynamical state of the cluster
\cite{52,53}. Such joint X-ray and weak-lensing studies, encompassing
the mass distribution out to the virial radius \cite{52} and
$r_{500}$ \cite{53}, clearly show a pronounced radial dependence
in the value of $f_{\rm gas}$ inferred from both weak-lensing
and the ratio of hydrostatic equilibrium mass to weak-lensing
mass. 

Ironically, the best-fit mean gas fractions that we have derived here
(for both $\Lambda$CDM and $R_{\rm h}=ct$) are comparable for the two
values of $\Delta$ used in the samples adopted in this paper, in spite
of the expected radial dependence in $f_{\rm gas}$. It appears that
these two effects, i.e., the radial dependence of $f_{\rm gas}$ and
the apparent breakdown of hydrostatic equilibrium, largely offset
each other, producing an almost constant mass fraction between
$\Delta=500$ and $\Delta=2,500$. Nonetheless, the model comparison 
we have carried out here will benefit considerably from the much 
more detailed and better spatially-resolved measurements that will
be made in the near future along the lines reported in refs.~\cite{52,53}.

In the long run, one would like to match the capabilities of techniques 
using Type Ia SNe, cluster number counts, weak lensing and BAO to study
the possible redshift dependence of $f_{\rm gas}$ and its implication for
cosmology. But in order to do this, one would need to measure $f_{\rm gas}$ 
to $\sim 5\%$ accuracy for large samples (i.e., $> 500$) of hot, massive 
clusters ($kT_e> 5$ keV), spanning the redshift range $0<z<2$ \cite{51}.
Though the Constellation-X Observatory and {\it XEUS}, for which these 
estimates were first developed, are no longer viable future missions, 
they have evolved into another possible project, known as The Advanced 
Telescope for High ENergy Astrophysics (ATHENA), which could contribute
to the resources necessary to carry out the required observations, if it 
ever reaches maturity.

\section*{Acknowledgments}
I am very grateful to Amherst College for its support through a John Woodruff
Simpson Lectureship, and to ISSI in Bern where some of this work was carried out.

\vskip 0.2in\noindent{\bf Ethics Statement.} This research poses no ethical considerations.

\vskip 0.2in\noindent{\bf Data Accessibility Statement.} All data used in this paper were
previously published, as indicated in Table 1.

\vskip 0.2in\noindent{\bf Competing Interests Statement.} We have no competing interests.

\vskip 0.2in\noindent{\bf Authors' contributions.} FM is the sole author of this paper.

\vskip 0.2in\noindent{\bf Funding.} Partial support for this work was provided by
the International Space Science Institute in Bern, Switzerland.

\end{document}